\documentclass[twoside,twocolumn,9pt]{article}
\usepackage{extsizes}
\usepackage[super,sort&compress,comma]{natbib} 
\usepackage[version=3]{mhchem}
\usepackage[left=1.5cm, right=1.5cm, top=1.785cm, bottom=2.0cm]{geometry}
\usepackage{balance}
\usepackage{mathptmx}
\usepackage{sectsty}
\usepackage{graphicx} 
\usepackage{lastpage}
\usepackage[format=plain,justification=justified,singlelinecheck=false,font={stretch=1.125,small,sf},labelfont=bf,labelsep=space]{caption}
\usepackage{float}
\usepackage{fancyhdr}
\usepackage{fnpos}
\usepackage[english]{babel}
\addto{\captionsenglish}{%
  \renewcommand{\refname}{Notes and references}
}
\usepackage{array}
\usepackage{droidsans}
\usepackage{charter}
\usepackage[T1]{fontenc}
\usepackage[usenames,dvipsnames]{xcolor}
\usepackage{setspace}
\usepackage[compact]{titlesec}
\usepackage{hyperref}
\usepackage{dcolumn}

\usepackage{epstopdf}

\usepackage{booktabs}

\definecolor{cream}{RGB}{222,217,201}

\begin{document}

\pagestyle{fancy}
\thispagestyle{plain}
\fancypagestyle{plain}{
\renewcommand{\headrulewidth}{0pt}
}

\makeFNbottom
\makeatletter
\renewcommand\LARGE{\@setfontsize\LARGE{15pt}{17}}
\renewcommand\Large{\@setfontsize\Large{12pt}{14}}
\renewcommand\large{\@setfontsize\large{10pt}{12}}
\renewcommand\footnotesize{\@setfontsize\footnotesize{7pt}{10}}
\makeatother

\renewcommand{\thefootnote}{\fnsymbol{footnote}}
\renewcommand\footnoterule{\vspace*{1pt}%
\color{cream}\hrule width 3.5in height 0.4pt \color{black}\vspace*{5pt}} 
\setcounter{secnumdepth}{5}

\makeatletter 
\renewcommand\@biblabel[1]{#1}            
\renewcommand\@makefntext[1]%
{\noindent\makebox[0pt][r]{\@thefnmark\,}#1}
\makeatother 
\renewcommand{\figurename}{\small{Fig.}~}
\sectionfont{\sffamily\Large}
\subsectionfont{\normalsize}
\subsubsectionfont{\bf}
\setstretch{1.125} 
\setlength{\skip\footins}{0.8cm}
\setlength{\footnotesep}{0.25cm}
\setlength{\jot}{10pt}
\titlespacing*{\section}{0pt}{4pt}{4pt}
\titlespacing*{\subsection}{0pt}{15pt}{1pt}

\fancyfoot{}
\fancyfoot[LO,RE]{\vspace{-7.1pt}\includegraphics[height=9pt]{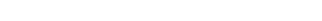}}
\fancyfoot[CO]{\vspace{-7.1pt}\hspace{13.2cm}\includegraphics{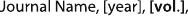}}
\fancyfoot[CE]{\vspace{-7.2pt}\hspace{-14.2cm}\includegraphics{head_foot/RF}}
\fancyfoot[RO]{\footnotesize{\sffamily{1--\pageref{LastPage} ~\textbar  \hspace{2pt}\thepage}}}
\fancyfoot[LE]{\footnotesize{\sffamily{\thepage~\textbar\hspace{3.45cm} 1--\pageref{LastPage}}}}
\fancyhead{}
\renewcommand{\headrulewidth}{0pt} 
\renewcommand{\footrulewidth}{0pt}
\setlength{\arrayrulewidth}{1pt}
\setlength{\columnsep}{6.5mm}
\setlength\bibsep{1pt}

\makeatletter 
\newlength{\figrulesep} 
\setlength{\figrulesep}{0.5\textfloatsep} 

\newcommand{\topfigrule}{\vspace*{-1pt}%
\noindent{\color{cream}\rule[-\figrulesep]{\columnwidth}{1.5pt}} }

\newcommand{\botfigrule}{\vspace*{-2pt}%
\noindent{\color{cream}\rule[\figrulesep]{\columnwidth}{1.5pt}} }

\newcommand{\dblfigrule}{\vspace*{-1pt}%
\noindent{\color{cream}\rule[-\figrulesep]{\textwidth}{1.5pt}} }

\makeatother

\twocolumn[
  \begin{@twocolumnfalse}
{\includegraphics[height=30pt]{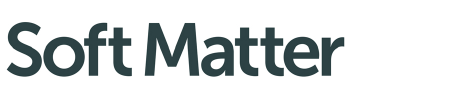}\hfill\raisebox{0pt}[0pt][0pt]{\includegraphics[height=55pt]{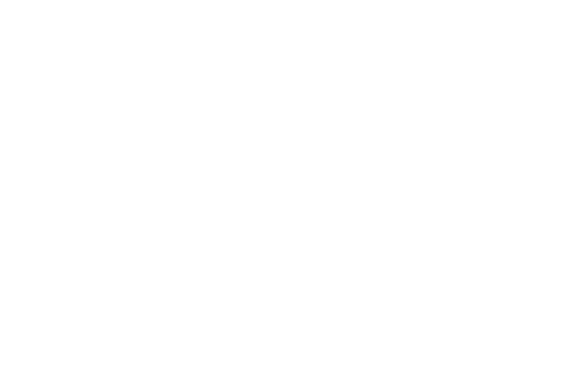}}\\[1ex]
\includegraphics[width=18.5cm]{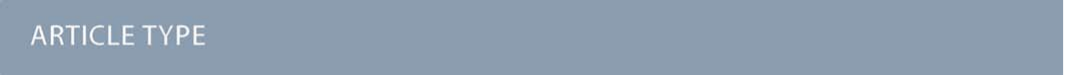}}\par
\vspace{1em}
\sffamily
\begin{tabular}{m{4.5cm} p{13.5cm} }

\includegraphics{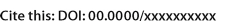} & \noindent\LARGE{\textbf{Sedimentation equilibrium as a probe of the pressure equation of state of active colloids}} \\
\vspace{0.3cm} & \vspace{0.3cm} \\

& \noindent\large{Yunhee Choi,$^\ddag$\textit{$^{a}$} Elijah Schiltz-Rouse,$^\ddag$\textit{$^{a}$} Parvin Bayati,$^\ddag$\textit{$^{a}$} and Stewart A. Mallory$^{\ast}$\textit{$^{a,b}$}} \\

\includegraphics{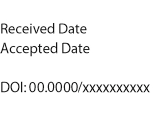} & \noindent\normalsize{We introduce a theoretical and computational framework for extracting the pressure equation of state (EoS) of an active suspension from its steady-state sedimentation profile.
As EoSs are prerequisites for many theories in active matter, determining how pressure depends on key parameters such as density, activity, and interparticle interactions is essential to make quantitative predictions relevant to materials design and engineering applications.
Focusing on the one-dimensional active Brownian particle (1D-ABP) model, we show that the pressure measured in a homogeneous periodic system can be recovered from the spatial profiles established in sedimentation equilibrium.
Our approach is based on exact mechanical considerations and provides a direct route for determining pressure from experimentally measurable quantities.
This work compares sedimentation-derived equations of state with those obtained from periodic simulations, establishing a foundation for using sedimentation as a generic tool to characterize the behavior of active suspensions.} \\

\end{tabular}

 \end{@twocolumnfalse} \vspace{0.6cm}

  ]

\renewcommand*\rmdefault{bch}\normalfont\upshape
\rmfamily
\section*{}
\vspace{-1cm}


\footnotetext{\textit{$^{a}$~Department of Chemistry, The Pennsylvania State University, University Park, Pennsylvania, 16802, USA; E-mail: sam7808@.psu.edu}}
\footnotetext{\textit{$^{b}$~Department of Chemical Engineering, The Pennsylvania State University, University Park, Pennsylvania, 16802, USA. }}


\footnotetext{\ddag~These authors contributed equally to this work.}





\begin{figure*}[h!]
\centering
\includegraphics[width=.95\textwidth]{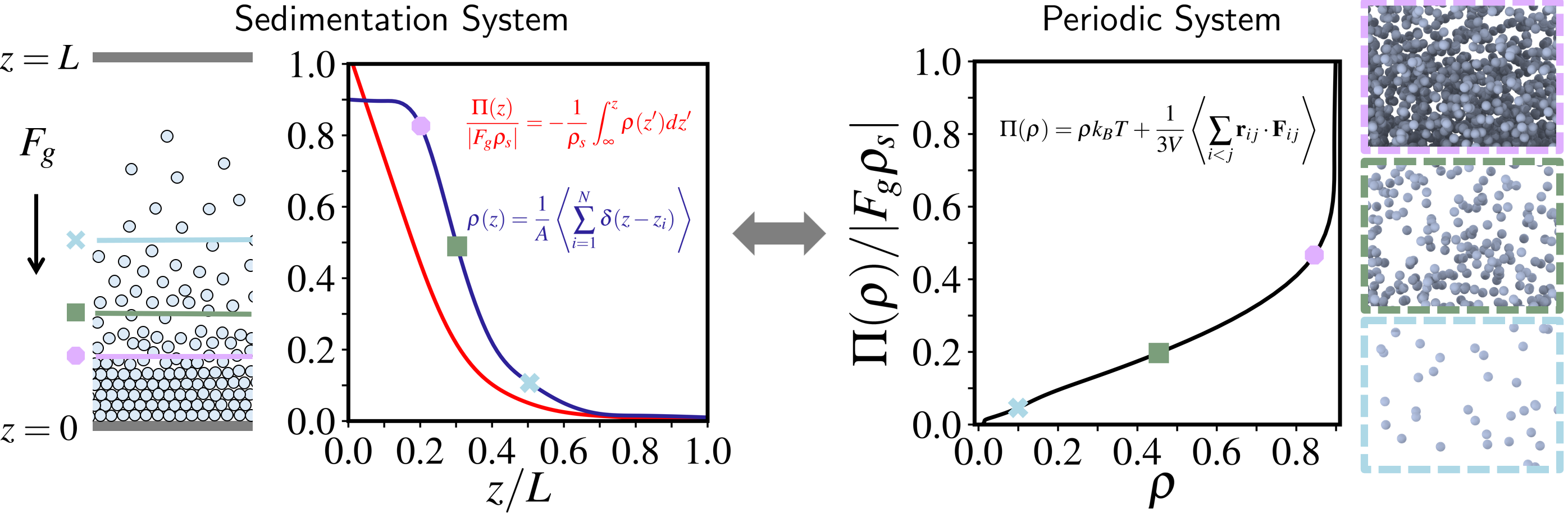}
\caption{\protect\small
{
Schematic for determining the equation of state (EoS) from the sedimentation profile of a passive colloidal suspension.  
\textbf{(left)} Spherical colloidal particles sediment under the presence of a gravitational force \( F_g \) in a container of height \( L \), establishing a steady-state inhomogeneous density profile. Colored horizontal lines denote reference heights associated with distinct local densities.  
\textbf{(middle)} The density profile \( \rho(z) \) (blue) decreases with height, and the corresponding pressure profile \( \Pi(z) \) (red) is obtained via integration of the force balance [eqn~(\ref{eq:7})].  
\textbf{(right)} Combining the density and pressure profiles yields the mechanical equation of state \( \Pi(\rho) \).  
Snapshots show periodic systems with densities matching those at the marked heights in the sedimentation system, demonstrating equivalence between local sedimentation pressure and bulk pressure in homogeneous periodic systems.
}}
\label{fig:1}
\end{figure*}

\section*{Introduction}

Sedimentation equilibrium is one of the most powerful and enduring methods for probing the physical properties of colloidal suspensions.~\cite{Perrin1910Brown,Piazza2014-xo}  
In a typical experimental setup, micron-sized particles sediment under gravity in a confined container, producing an inhomogeneous steady-state density profile.  
Equilibrium is achieved when osmotic pressure gradients exactly balance gravitational settling, allowing direct access to the system’s mechanical properties and interparticle interactions.~\cite{Batchelor1976Sedi, Pusey1986sedi}  

Historically, this approach played a central role in establishing the statistical thermodynamics of colloids.~\cite{Levine1934Sedi}
Perrin’s seminal work showed that the sedimentation profile of colloidal particles follows a Boltzmann distribution, enabling one of the first quantitative estimates of Avogadro’s number.~\cite{Perrin1910Brown} 
Ultimately, this early work gave irrefutable evidence for the atomic hypothesis and the discrete nature of matter.~\cite{perrin1916atoms}
Later, theoretical developments further revealed that sedimentation profiles encode the mechanical equation of state (EoS), linking local density to pressure.  
This insight enabled precise determination of the pressure EoS and facilitated characterization of interparticle interactions across a wide range of passive colloidal systems.~\cite{Barrat1992-vo, Biben1993-ri, Biben1994-ca, Martin1994-qz, Piazza1993-rq, rutgers1996measurement, buzzaccaro2007sticky, Hachisu1982-is}  
Together, these contributions underscore the versatility of sedimentation equilibrium as a fundamental tool in soft matter research.
However, extending this framework to active suspensions remains a significant challenge.  

Active colloids, including catalytic Janus particles,~\cite{paxton2004catalytic,howse2007self,palacci2014light,Bayati2024-fp, Bayati2016-bf,bechinger2016active} motile bacteria,~\cite{be2019statistical,maggi2013motility,mushenheim2014dynamic, schwarz2012phase,chen2015dynamic, aranson2022bacterial} and active droplets,~\cite{izri2014self,seemann2016self,jin2018chemotactic,maass2024self,Balaj2024-eq} are micron-size particles that can convert environmental or chemical energy into directed motion, violating the principle of detailed balance and thermodynamic equilibrium at a single particle level.
At a macroscopic and collective level, active suspensions evolve toward nonequilibrium steady states.  
As a result, standard tools of equilibrium statistical mechanics no longer apply. 
Nonetheless, equations of state have become central to theoretical studies of active matter, informing our understanding of motility-induced phase separation, anomalous transport, and collective dynamics.~\cite{Mallory2014-gd, Smallenburg2015-hy, Solon2015-lt, Marini-Bettolo-Marconi2015-tz, Solon2015-hu, Nikola2016-vw, Levis2017-cu,  Patch2017-ca, Jamali2018-fy, Duzgun2018-yg, Mallory2020-cx, Mallory2021-ux, Dulaney2021-me, Omar2021-gn, Speck2021-ax, Omar2023-kn} 
This is largely because pressure is a mechanical quantity that remains well-defined in nonequilibrium systems.~\cite{Takatori2014-cp, Winkler2015-qb, Omar2020-kr, speck2016ideal, joyeux2016pressure, marini2017pressure, epstein2019statistical, klamser2018thermodynamic}
In simulation, pressure is typically computed for homogeneous active systems using virial-like expressions incorporating interparticle forces and persistent self-propulsion.  
While the sedimentation of active colloids has received considerable attention,~\cite{Liu2024-ai, Wolff2013-zo, Palacci2010-ke, Scagliarini2022-uf, Chen2016-ev, Urena-Marcos2023-rd, Carrasco-Fadanelli2023-fe, Torres-Maldonado2022-wm, Singh2021-bz, Kuhr2017-be, Ginot2018-ig, Hermann2018-ar, Enculescu2011-wd, maldonado2024sedimentation, Wysocki2020-lm, Ginot2015-np, Ruhle2020-lc, Das2020-rz, Ruhle2018-np, vachier2019dynamics, Fins-Carreira2024-wb, van2025tunable, Kuhr2019-lb, Castonguay2023-ji} a quantitative link between sedimentation profiles and the pressure measured in bulk homogeneous systems has yet to be firmly established.

Focusing on the one-dimensional active Brownian particle (1D-ABP) model with purely repulsive interactions, we develop an exact theoretical framework that relates the sedimentation profile to the mechanical equation of state measured in periodic simulations. 
We identify the distinct contributions to the pressure from the local force balance and compare sedimentation-derived results to bulk simulations across a range of system parameters.  
This comparison serves as proof of concept for using sedimentation as a practical tool to extract pressure EoSs in active matter.  
Our findings lay the groundwork for extending this approach to higher-dimensional systems and experimental realizations.

\subsection*{Passive Sedimentation Equilibrium}

We begin by outlining the standard procedure for extracting the mechanical EoS of a passive colloidal suspension from its steady-state sedimentation profile.~\cite{Piazza2014-xo}  
This framework is illustrated schematically in Fig.~\ref{fig:1}, which shows how a measured local number density profile \( \rho(z) \), when combined with the appropriate force balance, yields the corresponding spatial profile of the osmotic pressure \( \Pi(z) \).  
Eliminating the vertical coordinate \( z \) between these two quantities allows one to obtain the homogeneous equation of state \( \Pi(\rho) \).  
The remainder of this section details this approach and demonstrates its equivalence to the virial route commonly used in periodic systems.

\begin{table*}[th!]
\centering
\caption{
Comparison of expressions for pressure components in sedimentation and periodic systems of passive colloidal suspensions. 
The table summarizes how total, Brownian, and collisional pressures are computed in each geometry. 
In sedimentation systems, pressure arises from integrating the density profile under gravity, whereas in periodic systems, pressure is determined from bulk properties using virial expressions.
}\renewcommand{\arraystretch}{2.0}
\begin{tabular}{@{} l >{\raggedright\arraybackslash}p{8.5cm} >{\raggedright\arraybackslash}p{6.5cm} @{}}
\toprule
 & \hspace{0.4cm}{Sedimentation System} & {Periodic System} \\
\midrule
Total Pressure 
& \hspace{0.4cm}\( \Pi(z') = F_g \displaystyle\int_{\infty}^{z'} \rho(z) \, \mathrm{d}z \) 
& \( \Pi = \rho k_B T + \dfrac{1}{3V} \left\langle \displaystyle \sum_{i<j} \mathbf{r}_{ij} \cdot \mathbf{F}_{ij} \right\rangle \) \\
\addlinespace[0.3em]
Brownian Pressure
& \hspace{0.4cm}\( \Pi_b(z) = \rho(z) k_B T \) 
& \( \Pi_b = \rho k_B T \) \\
\addlinespace[0.3em]
Collisional Pressure
& \hspace{0.4cm}\( \Pi_c(z') = -\rho(z') k_B T + F_g \displaystyle\int_{\infty}^{z'} \rho(z) \, \mathrm{d}z \) 
& \( \Pi_c = \dfrac{1}{3V} \left\langle \displaystyle \sum_{i<j} \mathbf{r}_{ij} \cdot \mathbf{F}_{ij} \right\rangle \) \\
\bottomrule
\end{tabular}
\label{tab:1}
\end{table*}

As shown in Fig.~\ref{fig:1}, we consider a system of \( N \) passive spherical colloidal particles suspended in a solvent at temperature \( T \), confined between two horizontal walls at \( z = 0 \) and \( z = L \).  
Each particle undergoes Brownian motion and interacts with other particles via an isotropic pair potential.  
In addition, all particles experience a weak gravitational force of strength $F_g$ acting in the negative \( z \)-direction.  
By symmetry, the system is translationally invariant in the \( xy \)-plane, and we define the local number density as
\begin{equation}
\rho(z) = \frac{1}{A} \left\langle \sum_{i=1}^N \delta(z - z_i) \right\rangle,
\label{eq:1}
\end{equation}
where \( A \) is the cross-sectional area perpendicular to the direction of gravity and \( z_i \) is the vertical position of particle \( i \).  
The density profile satisfies the normalization condition
\begin{equation}
\int_0^L \rho(z) \, \mathrm{d}z = \rho_s,
\label{eq:2}
\end{equation}
where \( \rho_s = N / A \) is the bulk cross-sectional number density.

Generally, the temporal and spatial evolution of the local number density \( \rho(z) \) is governed by coupled conservation equations for mass and momentum.  
Derivations of these governing equations for colloidal systems can be found in several standard texts.~\cite{Dhont1996-wj, Russel2012-ea, Graham2018-gx} 
Here, we focus on the steady-state behavior, where the density profile is time-independent and determined by the local mechanical force balance
\begin{equation}
\partial_z \sigma_{zz}(z) + b_z(z) = 0,
\label{eq:3}
\end{equation}
where \( \sigma_{zz}(z) \) is the \( zz \)-component of the stress tensor and \( b_z(z) \) is the \( z \)-component of the body force density.  
In simple terms, eqn~(\ref{eq:3}) states that external body forces due to gravity or confinement are locally balanced by osmotic stress gradients.

In the limit of weak inhomogeneity, where the density varies slowly with \( z \), a local density approximation is appropriate, and the stress can be expressed as \( \sigma_{zz}(z) = -\Pi(z) \), where \( \Pi(z) \) is the osmotic pressure in the \( z \)-direction.~\cite{Barrat1992-vo} 
For passive colloidal suspensions, the osmotic pressure consists of thermal and interaction contributions:
\begin{equation}
\Pi(z) = \Pi_b(z) + \Pi_c(z) = \rho(z) k_B T + \Pi_c(z),
\label{eq:4}
\end{equation}
where \( \Pi_b(z) = \rho(z) k_B T \) is the thermal (Brownian) pressure, and \( \Pi_c(z) \) is the interaction pressure.
The body force density accounts for gravity and confinement:
\begin{equation}
b_z(z) = F_g \, \rho(z) + F_w(z) \, \rho(z),
\label{eq:5}
\end{equation}
where \( F_w(z) \) is the force the confining walls exert.  
For short-ranged, strongly repulsive boundaries, the wall contribution can be approximated as
\begin{equation}
F_w(z) \, \rho(z) \approx p_{\text{bot}} \, \delta(z) - p_{\text{top}} \, \delta(z - L),
\label{eq:6}
\end{equation}
where \( p_{\text{bot}} \) and \( p_{\text{top}} \) are the mechanical pressures on the bottom and top walls, respectively.  
Substituting these expressions into the force balance [eqn~(\ref{eq:3})] yields
\begin{equation}
-\partial_z \Pi(z) + F_g \, \rho(z) + p_{\text{bot}} \, \delta(z) - p_{\text{top}} \, \delta(z - L) = 0.
\label{eq:7}
\end{equation}

\vspace{0.1cm}
Integrating eqn~(\ref{eq:7}) across the full domain \( z \in (-\infty, \infty) \) gives the well-known result for the total pressure difference between walls:
\begin{equation}
\Delta p = p_{\text{bot}} - p_{\text{top}} = -F_g \, \rho_s \, .
\label{eq:8}
\end{equation}
In most experimental settings, the container is tall enough that no particles accumulate at the top wall, allowing the simplification \( p_{\text{top}} = 0 \).
With knowledge of the density profile \( \rho(z) \), the pressure at any height \( 0 < z' < L \) can be obtained by integrating eqn~(\ref{eq:7}):
\begin{equation}
\Pi(z') = F_g \int_{\infty}^{z'} \rho(z) \, \mathrm{d}z.
\label{eq:9}
\end{equation}
The interaction contribution to the pressure then follows by substituting eqn~(\ref{eq:4}) into eqn~(\ref{eq:9}):
\begin{equation}
\Pi_c(z') = -\rho(z') k_B T + F_g \int_{\infty}^{z'} \rho(z) \, \mathrm{d}z.
\label{eq:10}
\end{equation}
The procedure outlined above (see Fig.~\ref{fig:1}) provides a direct method for extracting the pressure equation of state \( \Pi(\rho) \) from the steady-state density profiles of a sedimenting passive colloidal system.  

\vspace{-0.1cm}
It is useful to compare these results to the bulk pressure obtained for a homogeneous system of \( N \) particles in a volume \( V \) with periodic boundary conditions.  
In such systems, the total pressure is typically computed using the microscopic virial expression:~\cite {Hansen2014-mc, Frenkel2001-ms, Allen2017-yk, Thompson2009-bk}
\begin{align}
\Pi(\rho) &= \Pi_b(\rho) + \Pi_c(\rho) = \rho k_B T + \frac{1}{3 V} \left\langle \sum_{i=1}^N \mathbf{r}_{i} \cdot \mathbf{F}_{i} \right\rangle \nonumber \\
&= \rho k_B T + \frac{1}{3 V} \left\langle \sum_{i<j} \mathbf{r}_{ij} \cdot \mathbf{F}_{ij} \right\rangle,
\label{eq:11}
\end{align}
where \( \rho = N / V \) is the bulk number density, \( \mathbf{r}_{i} \) is the position of particle \( i \), and \( \mathbf{F}_{i} \) is the net force on particle \( i \) from all other particles.  
The first term represents the Brownian contribution, while the second term accounts for interparticle interactions.  
The second line follows from assuming pairwise additive forces, where \( \mathbf{r}_{ij} = \mathbf{r}_i - \mathbf{r}_j \) and \( \mathbf{F}_{ij} \) is the force exerted by particle \( j \) on particle \( i \).  
The various pressure components measured in a periodic system should agree with those extracted from the sedimentation profile under the local density approximation.  
For clarity, all relevant expressions for pressure are summarized in Table~\ref{tab:1}.

\section*{1D-ABP Model}

\begin{figure}[t!]
\centering
\includegraphics[width=.45\textwidth]{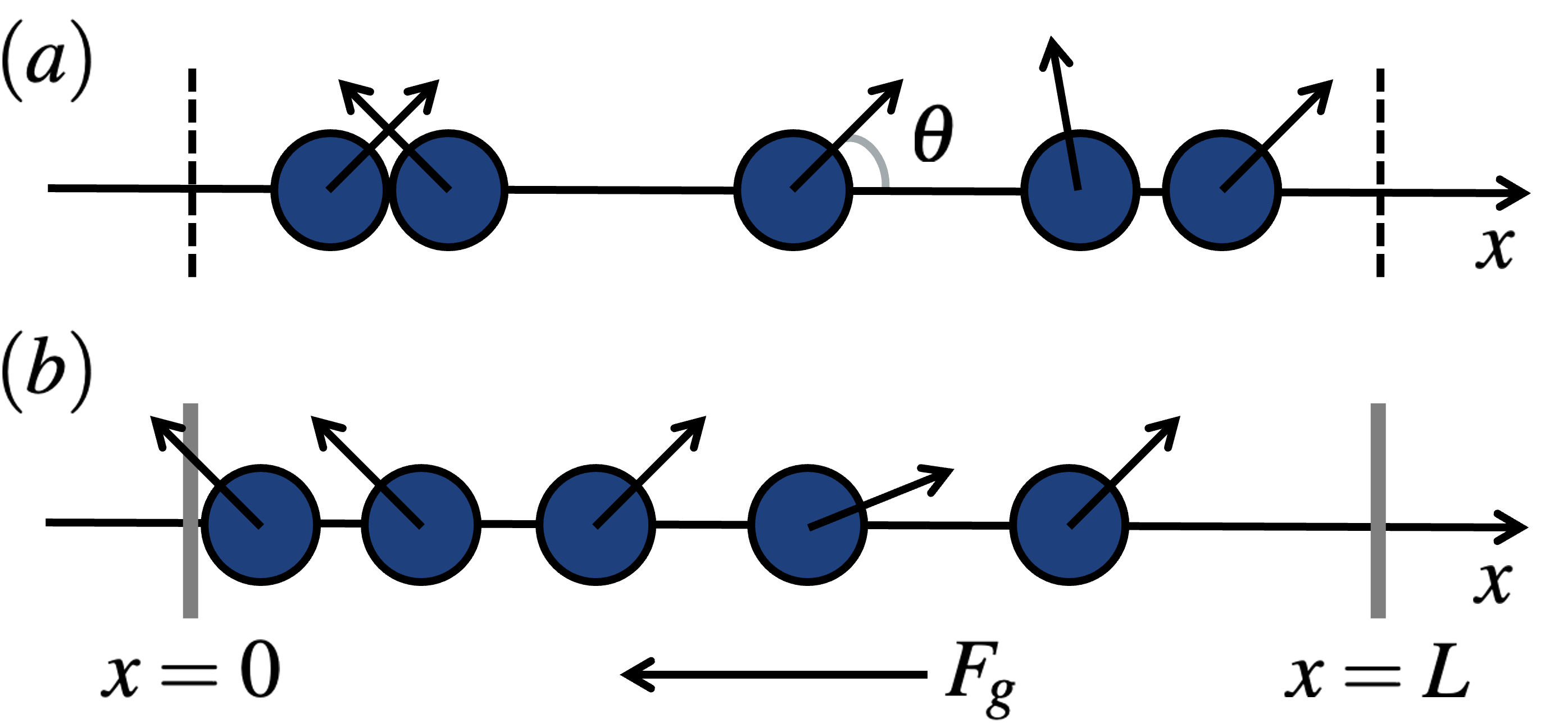}
\caption{\small Schematic of the one-dimensional active Brownian particle (1D-ABP) model studied in this work. In both geometries, particles are confined to a narrow channel that enforces single-file motion and experience a self-propulsion force \( F_a = \gamma U_a \cos \theta \) along their orientation. (a) In the periodic system, particles move on a closed loop without external potential. (b) In the sedimentation geometry, particles are confined by hard walls and subject to a constant gravitational force \( F_g \).}
\label{fig:2}
\end{figure}

To investigate if a similar procedure can be applied to active colloids, we consider two variants of a one-dimensional active Brownian particle (1D-ABP) model, illustrated in Fig.~\ref{fig:2}.  
This minimal model captures the essential features of self-propulsion and interparticle interactions while remaining analytically and computationally tractable.  
Moreover, its phase behavior and phenomenology are well characterized in prior studies.~\cite{Schiltz-Rouse2023-wg,Akintunde2025-cs}
Although we focus on this 1D geometry, the theoretical framework presented is readily extendable to higher-dimensional active systems.

In both system variants, \( N \) active Brownian disks are confined to a narrow channel of length \( L \).  
The channel width is small enough to prevent particle overtaking, enforcing single-file motion and effectively reducing the dynamics to one dimension.  
The first variant [Fig.~\ref{fig:2}a] employs periodic boundary conditions and represents a homogeneous, bulk system.  
The second variant [Fig.~\ref{fig:2}b] consists of a finite channel bounded by rigid walls, with particles subject to a constant gravitational force \( F_g \) acting in the negative \( x \)-direction.

In both geometries, each particle experiences a self-propulsion force \( F_a = \gamma U_a \cos \theta \), where \( \gamma \) is the translational drag coefficient, \( U_a \) is the propulsion speed, and \( \theta \) is the angle between the particle’s orientation and the \(x\)-axis.  
The orientation undergoes rotational diffusion in the plane perpendicular to the channel, with a characteristic reorientation time \( \tau_R \).  
Neglecting translational Brownian motion and hydrodynamic interactions, the overdamped Langevin equations governing the particle dynamics are
\begin{subequations}
\label{eq:12}
\begin{align}
\dot{x} &= U_a \cos \theta + \frac{1}{\gamma} \left( F_c + F_g + F_w \right), \\
\dot{\theta} &= \xi(t),
\end{align}
\end{subequations}
where \( F_c \) is the interparticle collision force, \( F_w \) is the force from the channel walls, and \( F_g \) is the gravitational force.  
Both \( F_c \) and \( F_w \) are derived from a short-ranged, purely repulsive Weeks–Chandler–Andersen (WCA) potential~\cite{Weeks1971-qh} with interaction strength \( \varepsilon \) and Lennard-Jones diameter \( \sigma \).  
We choose \( \varepsilon / (F_a \sigma) = 100 \) to approximate hard-particle behavior.  
The rotational noise \( \xi(t) \) is characterized by  \( \langle \xi(t) \rangle  = 0\) and autocorrelation \( \langle \xi(t) \xi(t') \rangle = (2 / \tau_R)\delta(t - t') \).  
For the periodic system, \( F_g = F_w = 0 \).

Initial configurations are generated by randomly placing particles in the channel without overlap, with orientation angles drawn from a uniform distribution.  
Simulations were performed using \texttt{HOOMD-blue}~\cite{Anderson2020-wl} with \( N = 100 \) particles in the sedimentation geometry and \( N = 1000 \) in the periodic geometry, using a timestep \( \delta t = 10^{-5} \) and trajectory lengths of at least \( 4 \times 10^8 \) steps. 

\begin{table*}[ht!]
\centering
\caption{
Comparison of pressure component expressions in active sedimentation and periodic systems.
The table summarizes how the total, swim, and collisional pressure are computed in sedimenting and homogeneous active Brownian particle (ABP) systems.  
In the sedimentation system, pressure components are derived from spatial profiles of number density and polar order via a mechanical force balance.  
In the periodic system, pressure is computed from virial-like expressions involving particle positions and forces.     
}\renewcommand{\arraystretch}{2.0}
\begin{tabular}{@{} l >{\raggedright\arraybackslash}p{8.5cm} >{\raggedright\arraybackslash}p{6.5cm} @{}}
\toprule
 & \hspace{0.4cm}{Sedimentation System} & {Periodic System} \\
\midrule
Total Pressure 
& \hspace{0.4cm}\( P(x') = F_g \displaystyle\int_{\infty}^{x'} \rho(x) \, \mathrm{d}x \) 
& \( P = \rho \langle x F_{\text{net}} \rangle = \rho \langle x F_c \rangle + \rho \langle x F_a \rangle \) \\
\addlinespace[0.3em]
Swim Pressure
& \hspace{0.4cm}\( P_s(x') = \rho(x') \langle v(x') F_a(x') \rangle \tau_R =  -\gamma U_a \displaystyle \int_{\infty}^{x'} m(x) \, dx \) 
& \( P_s = \rho \langle x F_a \rangle = \rho \langle v F_a \rangle \tau_R = \rho \gamma \langle v^2 \rangle \tau_R  \) \\
\addlinespace[0.3em]
Collisional Pressure
& \hspace{0.4cm}\(
\begin{aligned}
P_c(x') &= \gamma U_a \displaystyle \int_{\infty}^{x'} m(x) \, dx + F_g \displaystyle \int_{\infty}^{x'} \rho(x) \, dx \\
P_c(x') &= -P_s(x') + F_g \displaystyle \int_{\infty}^{x'} \rho(x) \, dx
\end{aligned}
\)
& \( P_c = \rho \langle x F_c \rangle = \rho \left\langle \displaystyle\sum_{i < j} x_{ij} F_{ij} \right\rangle \) \\
\bottomrule
\end{tabular}
\label{tab:2}
\end{table*}

\subsection*{Phenomenology \& Pressure of Periodic 1D-ABPs}

The phenomenology and pressure behavior of the periodic 1D-ABP model have been examined in detail in an earlier study.~\cite{Akintunde2025-cs, Schiltz-Rouse2023-wg}  
For the interested reader, these studies analysed this model's clustering behavior, pressure, and transport properties. 
To summarize, the system is governed by two key dimensionless parameters: the packing fraction \( \phi = \rho \sigma_p \), where \( \rho = N/L \) is the line density and \( \sigma_p = 2^{1/6} \sigma \) is the effective particle diameter, and the active Péclet number \( \mathrm{Pe} = \ell_0/\sigma \), where \( \ell_0 = U_a \tau_R \) is the run length of an isolated particle.  
The Péclet number quantifies the persistence of directed motion relative to particle size.  
In the limit \( \mathrm{Pe} \ll 1 \), the dynamics reduce to those of the equilibrium Tonks gas.~\cite{Tonks1936-pf, Helfand1961-gx, Sells1953-ft, Barnes1999-qq, Bishop1974-qc}
At larger values of \( \mathrm{Pe} \), persistent motion leads to large dynamic clusters, although the system does not undergo motility-induced phase separation.~\cite{Schiltz-Rouse2023-wg,Dolai2020-cu,Gutierrez2021-mu,Mukherjee2023-ix}  
Notably, the periodic 1D-ABP system remains homogeneous and isotropic for all values of \( \phi \) and \( \mathrm{Pe} \).

In the periodic 1D-ABP model, the total pressure \( P \) is defined via the micromechanical virial~\cite{Takatori2014-cp} as
\begin{equation}
P = \rho \langle x F_{\text{net}} \rangle = \rho \langle x F_c \rangle + \rho \langle x F_a \rangle \,,
\label{eq:13}
\end{equation}
where \( F_{\text{net}} = F_a + F_c \) is the net force acting on a particle.  
This expression is the active analog of the virial pressure used for passive systems [eqn~(\ref{eq:11})].
The first term on the right-hand side defines the collisional pressure:
\begin{equation}
P_c = \rho \langle x F_c \rangle = \frac{\rho}{2} \left\langle \sum_{i \ne j} x_{ij} F_{ij} \right\rangle \,,
\label{eq:14}
\end{equation}
where \( F_{ij} \) is the pairwise force between particles \( i \) and \( j \), and \( x_{ij} \) is their separation.  
The last term of eqn~(\ref{eq:14}) assumes pairwise additive interactions.

The second term defines the swim pressure, which captures the contribution from self-propulsion:
\begin{equation}
P_s = \rho \langle x F_a \rangle = \rho \langle v F_a \rangle \tau_R = \rho \gamma \langle v^2 \rangle \tau_R \,.
\label{eq:15}
\end{equation}
The form of the swim pressure \(P_s = \rho \langle v F_a \rangle \tau_R\) is known as the active impulse representation,~\cite{Patch2017-ca,Solon2018-tk} and the final equivalent form of the swim pressure is \( P_s = \rho \gamma \langle v^2 \rangle \tau_R \), which highlights an analogy to the Brownian pressure, with the swim pressure being proportional to the mean-square velocity.~\cite{Schiltz-Rouse2023-wg} 
However, unlike thermal systems, \( \langle v^2 \rangle \) depends nontrivially on both \( \mathrm{Pe} \) and \( \phi \).  
These two expressions for the swim pressure are valid for active systems with a well-defined reorientation time \( \tau_R \) and no interparticle torques—conditions satisfied by the ABP model.

In the absence of interparticle interactions (\( F_c = 0 \)), the pressure reduces to the ideal expression
\begin{equation}
P_0 = \frac{1}{2} \rho \gamma U_a^2 \tau_R \,.
\label{eq:16}
\end{equation}

\vspace{-0.23cm}
As part of our prior work,~\cite{Schiltz-Rouse2023-wg} we showed that the reduced swim pressure \( \mathcal{P}_s = P_s / P_0 \) is directly related to a dimensionless kinetic temperature \( \mathcal{T}_k \), which captures the suppression of particle speed due to collisions:
\begin{equation}
\mathcal{P}_s = \frac{2 \langle v F_a \rangle}{\gamma U_a^2} = \frac{2 \langle v^2 \rangle}{U_a^2} = 1 - \frac{2 \langle F_c^2 \rangle}{(\gamma U_a)^2} = \mathcal{T}_k \,.
\label{eq:17}
\end{equation}
Using scaling arguments based on the statistics of interparticle collisions, we derived a quantitative analytical expression for \( \mathcal{P}_s \) as a function of packing fraction \( \phi \) and Péclet number \( \text{Pe} \):
\begin{equation}
\mathcal{P}_s = \frac{1}{9b^2}\left[2 \cos \left(\frac{1}{3}\arccos\left(\frac{27}{2} b^2 - 1\right)\right) - 1\right]^2 \,,
\label{eq:18}
\end{equation}
where \( b = \alpha \text{Pe} \phi /(1-\phi) \), with \( \alpha = c/(1 + \text{Pe})^d \), \( c = 1.1 \), and \( d = 0.05 \).
Furthermore, we also obtained simple expressions for the reduced collisional and total pressure:
\begin{align}
\mathcal{P}_c &= \mathcal{P}_s \left[\frac{\phi}{1 - \phi} \right], 
\label{eq:19} \\
\mathcal{P} &= \mathcal{P}_s \left[\frac{1}{1 - \phi} \right].
\label{eq:20}
\end{align}
Together, Eqs.~(\ref{eq:17})–(\ref{eq:20}) establish a direct and quantitative link between activity, packing fraction, and pressure in the 1D-ABP model.  
These results serve as a theoretical reference for evaluating sedimentation-based measurements of pressure.

\subsection*{Active Sedimentation Equilibrium}
We now outline a theoretical framework that connects the sedimentation profile of active Brownian particles (ABPs) to the pressure EoS established in periodic systems.  
Our analysis focuses on three central observables: the local number density, the local polar order, and the wall pressure.  
The local number density at position \( x \) is defined as
\begin{equation}
\rho(x) = \left\langle \sum_{i=1}^N \delta(x - x_i) \right\rangle \,,
\label{eq:21}
\end{equation}
and satisfies the normalization condition
\begin{equation}
\int_0^L \rho(x) \, dx = N \,.
\label{eq:22}
\end{equation}
The local polar order field characterizes spatial variations in particle orientation and is given by
\begin{equation}
m(x) = \left\langle \sum_{i=1}^N \cos(\theta_i) \, \delta(x - x_i) \right\rangle \,.
\label{eq:23}
\end{equation}
As there are no aligning torques in the ABP model, the total polar order must vanish at steady state:~\cite{Hermann2020-br}
\begin{equation}
M = \int_{-\infty}^{\infty} m(x) \, dx = 0 \,.
\label{eq:24}
\end{equation}

The governing continuum equations for ABPs have been derived in several prior studies.~\cite{Omar2023-kn,Saintillan2013-nq,epstein2019statistical}
Here, we present an abbreviated derivation tailored to the 1D-ABP system.  
At steady state, the system satisfies the local mechanical force balance:
\begin{equation}
\partial_x \sigma_{xx}(x) + b_x(x) = 0 \,,
\label{eq:25}
\end{equation}
where \( \sigma_{xx}(x) \) is the \( xx \)-component of the stress tensor and \( b_x(x) \) is the body force density in the \(x\)-direction.  
For 1D-ABPs under gravity, the body force density has three contributions—from self-propulsion, gravity, and wall interactions:
\begin{equation}
b_x(x) = \gamma U_a m(x) + F_g \rho(x) + F_w(x) \rho(x) \,.
\label{eq:26}
\end{equation}
Assuming a slowly varying density profile, we adopt a local density approximation and identify \( \sigma_{xx}(x) = -P_c(x) \), where \( P_c(x) \) is the collisional pressure.  
For strongly repulsive, short-ranged wall interactions, the generic force balance eqn~(\ref{eq:25}) becomes:
\begin{equation}
- \partial_x P_c(x) + \gamma U_a m(x) + F_g \rho(x) + p_{\text{bot}} \delta(x) - p_{\text{top}} \delta(x - L) = 0 \,,
\label{eq:27}
\end{equation}
with \( p_{\text{bot}} \) and \( p_{\text{top}} \) denoting the wall pressures at \( x = 0 \) and \( x = L \).  

Integrating eqn~(\ref{eq:27}) across the full spatial domain \(x \in (-\infty, \infty)\) and applying the global constraint on polar order [eqn~(\ref{eq:24})], the difference in wall pressure is given by
\begin{equation}
\Delta p = p_{\text{bot}} - p_{\text{top}} = -F_g N \,.
\label{eq:28}
\end{equation}
This result is analogous to the wall pressure difference obtained for passive systems [eqn~(\ref{eq:8})] and, notably, is independent of the form of the wall potential as discussed in prior work.~\cite{Solon2015-hu,Hermann2020-br}

Without loss of generality, we assume \( p_{\text{top}} = 0 \) and obtain the collisional pressure at position \( x' \) by integrating eqn~(\ref{eq:27}):
\begin{equation}
P_c(x') = \gamma U_a \int_{\infty}^{x'} m(x) \, dx + F_g \int_{\infty}^{x'} \rho(x) \, dx \,.
\label{eq:29}
\end{equation}
We can now compare the expression for collisional pressure for the active system [eqn~(\ref{eq:29})] with the passive case [eqn~(\ref{eq:10})].
Both expressions include a gravitational term involving the integral of the density profile.  
However, in passive Brownian systems, the thermal pressure appears explicitly and must be subtracted to isolate the interaction pressure.  
In active systems, by contrast, the pressure depends nonlocally on the orientation field through the integral of \( m(x) \).  
While the concept of swim pressure plays a central role in periodic geometries, it does not appear explicitly in the sedimentation force balance, underscoring a key distinction between passive and active systems.

\begin{figure}[t!]
\centering
\includegraphics[width=0.45\textwidth]{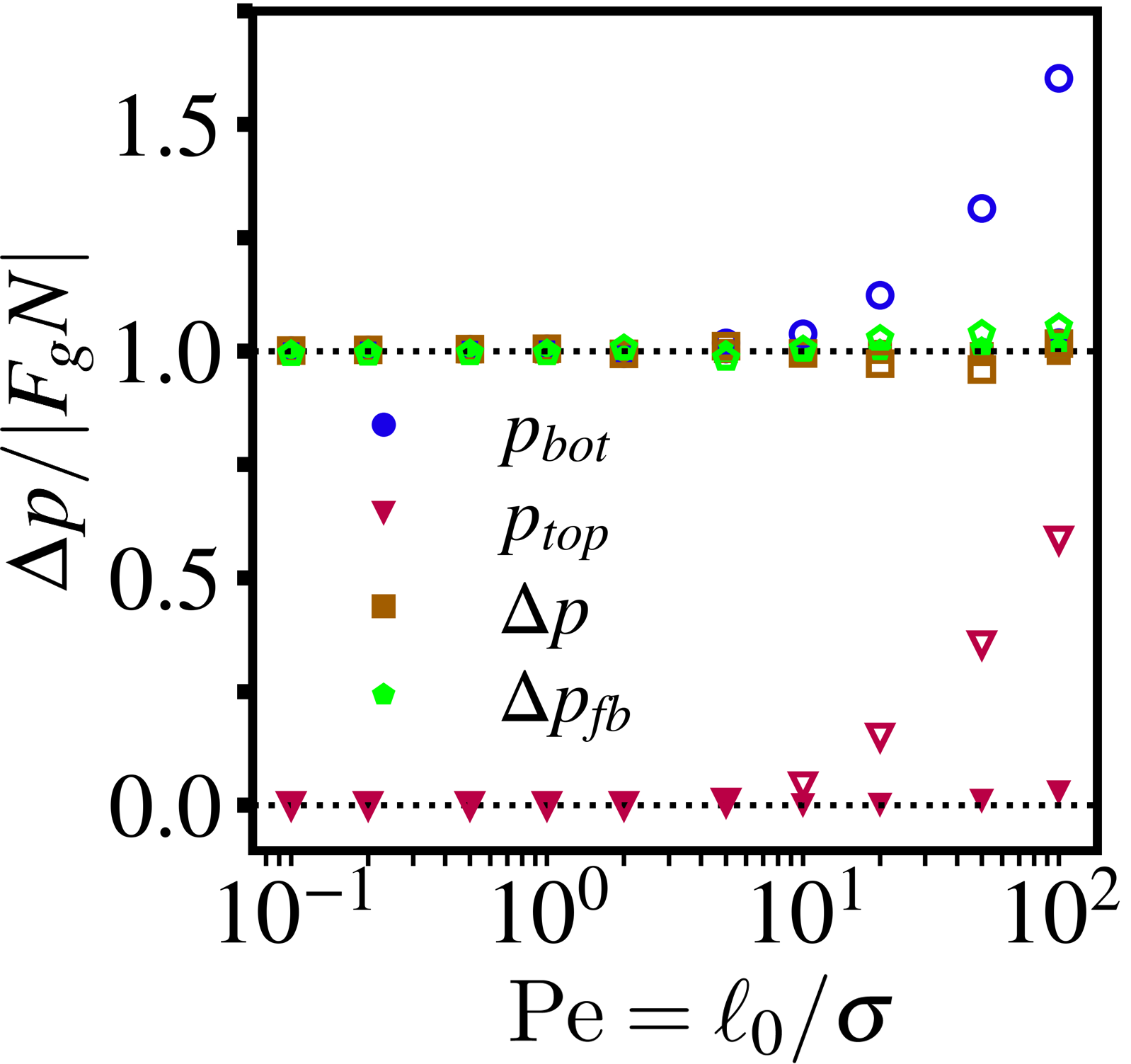}
\caption{\protect\small{
Normalized wall pressures in the sedimenting 1D-ABP system as a function of Péclet number. The bottom wall pressure (blue circles), top wall pressure (red triangles), pressure difference between the bottom and top walls from the simulation virials (beige squares), and the pressure difference between the top and bottom wall from the force balance (green pentagons) are shown for two gravitational forces: \(F_g = -0.05\) (filled markers) and \(F_g = -0.01\) (unfilled markers).  
All pressures are normalized by the theoretical prediction \(|F_g N|\).  
The normalized pressure difference remains close to unity across all conditions, confirming the force balance prediction.  
At high Péclet numbers, particles reach the top wall more frequently, resulting in a measurable pressure at the upper boundary.
}}
\label{fig:3}
\end{figure}

To explicitly introduce the swim pressure into the force balance [eqn~(\ref{eq:27})], we note that the polar order field \( m(x) \) can be expressed as the gradient of a polarization flux.  
This relationship arises naturally from coarse-graining the \( N \)-body Fokker–Planck equation, which yields a hierarchy of equations for the orientation moments.  
For 1D-ABPs, the steady-state conservation law for polar order is: ~\cite{Omar2023-kn,Saintillan2013-nq}
\begin{equation}
\gamma U_a m(x) = -\tau_R \, \partial_x j_{xx}^m(x) \,,
\label{eq:30}
\end{equation}
where the polarization flux is defined as
\begin{equation}
j_{xx}^m(x) = \left\langle \sum_i v_i F_{a,i} \, \delta(x - x_i) \right\rangle =  \rho(x)\langle v(x) F_a(x) \rangle \,.
\label{eq:31}
\end{equation}
Comparing Eqs.~(\ref{eq:30}) and (\ref{eq:31}) with the periodic system expression for swim pressure [eqn~(\ref{eq:15})], we can, by inspection, identify the polar order as the gradient of a local swim pressure:
\begin{equation}
\gamma U_a m(x) = - \partial_x P_s(x) = -\partial_x \left[\rho(x)\langle v(x) F_a(x) \rangle \tau_R \right] \,.
\label{eq:32}
\end{equation}
Here, we note that the local swim pressure can also be expressed in terms of the local mean-square velocity as \(P_s(x) = -\partial_x \left[\gamma\rho(x)\langle v(x)^2 \rangle \tau_R \right]\), which follows from the equivalent representations of swim pressure in periodic systems previously discussed. 
However, throughout this study, we express the local swim pressure in terms of the active impulse representation for brevity.

\begin{figure*}[ht!]
\centering
\includegraphics[width=0.95\linewidth]{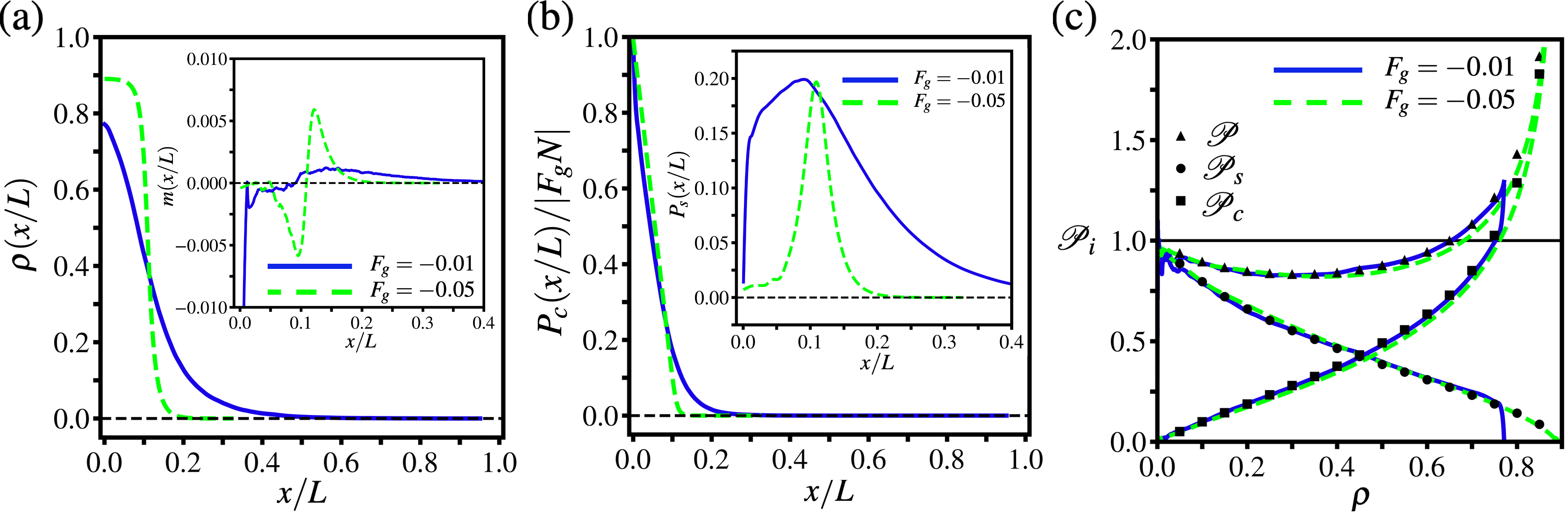}
\caption{\small{
Spatial profiles and pressure components for the 1D-ABP system at fixed Péclet number \(\mathrm{Pe} = 2\) and two gravitational forces \(F_g = -0.01\) (solid blue) and \(F_g = -0.05\) (dashed green). 
(a) Density profiles \(\rho(x)\), with inset showing the corresponding polar order \(m(x)\). 
(b) Collisional pressure \(P_c(x)\), normalized by \(|F_g N|\), with inset showing the swim pressure \(P_s(x)\). 
(c) Reduced total pressure \(\mathcal{P} \), swim pressure \(\mathcal{P}_s\), and collisional pressure \(\mathcal{P}_c\) as functions of density \(\rho\), comparing sedimentation-derived values (lines) with measurements in periodic systems (symbols). 
}}
\label{fig:4}
\end{figure*}

From eqn~(\ref{eq:32}), the local swim pressure at location \(x'\) can be determined by direct integration:
\begin{equation}
P_s(x') = \rho(x') \langle v(x') F_a(x') \rangle \tau_R = -\gamma U_a \int_{\infty}^{x'} m(x) \, dx \,.
\label{eq:33}
\end{equation}
Now, the collisional pressure [eqn~(\ref{eq:29})] can be rewritten in terms of the swim pressure by substituting eqn~(\ref{eq:33}):
\begin{equation}
P_c(x') = - P_s(x') + F_g \int_{\infty}^{x'} \rho(x) \, dx \,.
\label{eq:34}
\end{equation}
This leads to a total pressure defined as
\begin{equation}
P(x) = P_c(x) + P_s(x) \,,
\label{eq:35}
\end{equation}
which simplifies the force balance equation to
\begin{equation}
- \partial_x P(x) + F_g \rho(x) + p_{\text{bot}} \delta(x) - p_{\text{top}} \delta(x - L) = 0 \,.
\label{eq:36}
\end{equation}
Assuming again \( p_{\text{top}} = 0 \), the total pressure at position \( x' \) becomes
\begin{equation}
P(x') = F_g \int_{\infty}^{x'} \rho(x) \, dx \,.
\label{eq:37}
\end{equation}

\vspace{-0.1cm}
This result mirrors the passive sedimentation case [eqn~(\ref{eq:9})], indicating that—at the level of total pressure—the active system satisfies a formally similar balance.
Yet a key distinction remains: the total pressure includes a nonlocal swim component.  
Unlike Brownian or collisional pressures, which act locally through stress gradients, the swim pressure reflects the system’s orientational structure and is inherently nonlocal.~\cite{Omar2020-kr, Omar2023-kn}
Equation~(\ref{eq:33}) demonstrates that computing the swim pressure at \( x' \) requires either the full polar order field for \( x > x' \) or knowledge of the local density \(\rho(x')\) and velocity-active force correlation \(\langle v(x') F_a(x') \rangle\). 
Ultimately, the swim pressure or polar order contribution behaves more like a body force than a conventional stress. 
While eqn~(\ref{eq:37}) represents the total pressure at a virtual location \( x' \), it does not necessarily equal the pressure that would be exerted on a physical wall placed at that position.
As the introduction of a new object into the system will further modify the polar order field, and effectively the pressure at that location.
This subtlety arises as swim pressure reflects behavior across a finite spatial domain.

The above framework connects the spatial profiles of density and polarization to their corresponding pressure components in active systems.  
A summary of key expressions and their periodic counterparts is provided in Table~\ref{tab:2}.  

\section*{Results \& Discussion}

Next, we compare these results with simulations of the 1D-ABP model in periodic and sedimentation geometries.
As a first check of our theoretical framework, we verify the prediction of eqn~\ref{eq:28} for the pressure difference between the two walls, \(\Delta p = -F_g N\).  
To test this relation, we simulate the sedimenting 1D-ABP system under two gravitational forces, \(F_g = -0.01\) (open symbols) and \(F_g = -0.05\) (closed symbols), across a broad range of Péclet numbers.  
The wall pressures are computed by time-averaging the forces exerted on the walls, and we plot in Fig.~\ref{fig:3} the bottom and top wall pressures, as well as their difference, normalized by the theoretical value \(|F_g N|\).  
Across all conditions, the normalized pressure difference remains close to unity, in excellent agreement with our theoretical prediction.

At low Péclet numbers, particles are more susceptible to gravitational forces, leading to accumulation near the bottom wall and negligible pressure at the top boundary.  
As \(\mathrm{Pe}\) increases, particles gain sufficient persistence to explore more of the system, resulting in a growing top wall pressure.  
This accumulation is more pronounced for the weaker gravitational force (\(F_g = -0.01\)), where the run length better competes with gravity and redistributes particles toward the upper boundary.  
For the stronger gravitational force, the top wall pressure increases only at the highest Péclet numbers considered.

As an additional numerical check, we also compute the pressure difference from the measured density and polarization profiles via the integrated force balance:
\begin{equation}
\Delta p_{\text{fb}} = - \gamma U_a \int_{-\infty}^{\infty} m(x) \, dx - F_g \int_{-\infty}^{\infty} \rho(x) \, dx \,.
\label{eq:38}
\end{equation}
As shown in Fig.~\ref{fig:3} by the pentagonal symbols, this alternative approach provides an independent check and shows excellent agreement with the wall pressure measurements.

To illustrate how the EoS can be determined from sedimentation data, we examine in detail a representative case in Fig.~\ref{fig:4}.  
We focus on the 1D-ABP model at fixed activity (\(\mathrm{Pe} = 2\)) and consider two gravitational forces, \(F_g = -0.01\) and \(F_g = -0.05\), chosen to highlight how varying sedimentation profiles yield local pressure information.  
This example serves to demonstrate the whole procedure used throughout the study.

Figure~\ref{fig:4}a shows the steady-state number density \(\rho(x)\) and polar order \(m(x)\), computed from simulations using regularly spaced histogram bins of size (\(1.0\sigma_p\)) and post-processed using a Savitzky–Golay filter (window length 21, polynomial order 2) to smooth the profile.  
For both values of \(F_g\), \(\rho(x)\) decays monotonically with height, while stronger gravity leads to a sharp sedimentation layer near the bottom wall.
As expected, the density profile becomes broader for weaker gravitational forces, reflecting the enhanced persistence of active particles.  

The inset displays the polar order \(m(x)\), which integrates to zero within numerical precision, consistent with eqn~(\ref{eq:24}).  
In both cases, particles near the bottom wall align with the surface, yielding negative \(m(x)\).  
At weaker gravity, the polar order profile shows a pronounced dip near the wall (reaching \(-0.03\)), gradually becomes positive, and at longer distances decays to zero. 
For clarity, we have cropped this significant negative spike in polar order in Fig.~\ref{fig:4}a to better visualize the entire polar order profile. 
In contrast, for stronger gravity, \(m(x)\) is nearly zero at the wall due to crowding, becomes negative just above it, and transitions sharply to positive at the edge of the sedimentation layer.  
These contrasting profiles underscore how sedimentation strength controls structural and orientational organization in active systems.

Figure~\ref{fig:4}b shows the spatially resolved collisional pressure \(P_c(x)\), computed from eqn~(\ref{eq:29}) and normalized by \(|F_g N|\).  
At the bottom wall, \(P_c(x)\) converges to the expected mechanical pressure, validating consistency with the force balance.  
The pressure decays monotonically with height, mirroring the density profile, as expected for a system governed by short-range repulsive interactions. 
The inset shows the corresponding local swim pressure \(P_s(x)\), obtained from eqn~(\ref{eq:33}).  
This quantity exhibits nonmonotonic behavior: it increases with height, peaks near the crossover in polar order, and then decays to zero at the top wall.  

Theoretically, \(P_s(x)\) can be interpreted as the cumulative integral of the polar order or as a local polarization flux.  
This duality highlights the unique nature of the swim pressure---it reflects both global orientational structure and local particle dynamics.  
In the integral view, the value of \(P_s(x)\) at a given point depends nonlocally on the polar order across the system.  
In the flux-based interpretation, \(P_s(x)\) becomes a local quantity, expressed via eqn~(\ref{eq:33}) as \( \rho(x) \langle v(x) F_a(x) \rangle \tau_R \), where the correlation \( \langle v(x) F_a(x) \rangle \) captures suppression of motion due to collisions.  
While the swim pressure grows linearly with local density, this increase is tempered by the reduction in effective speed.  
As a result, \(P_s(x)\) reaches a maximum in intermediate-density regions where both density and activity are appreciable.

\begin{figure*}[ht!]
\centering
\includegraphics[width=0.95\textwidth]{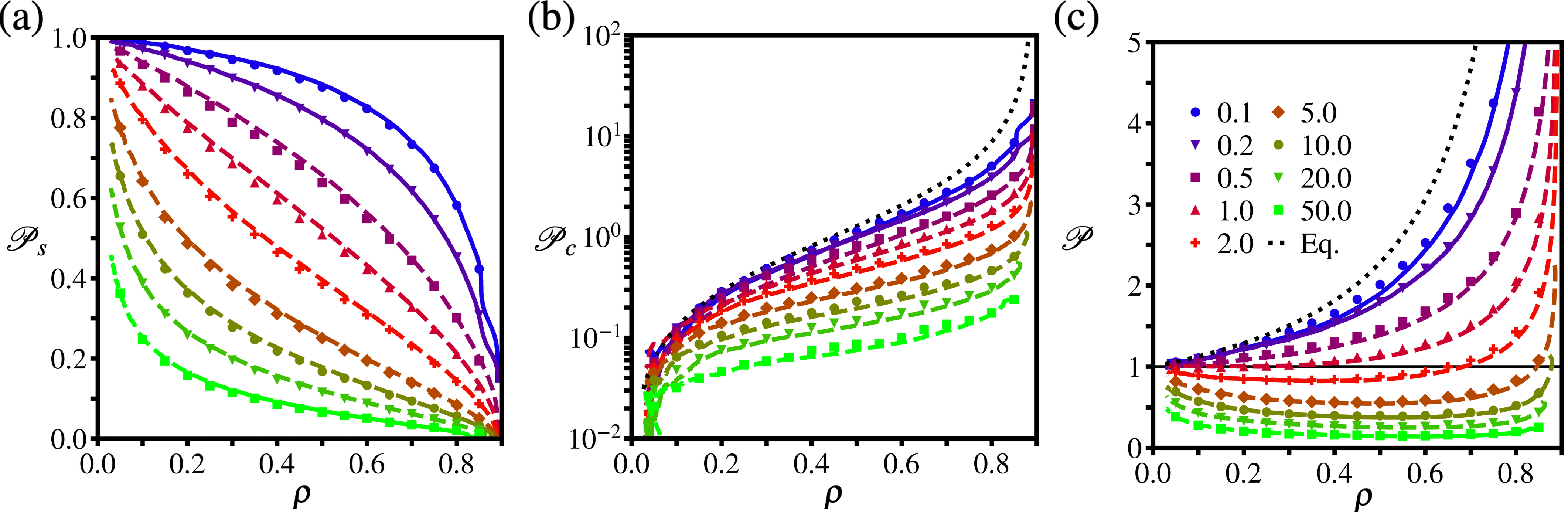}
\caption{\small{
Mechanical equation of state for the 1D-ABP system across a range of Péclet numbers. 
Panels show the (a) reduced swim pressure \( \mathcal{P}_s \), (b) collisional pressure \( \mathcal{P}_c \), and (c) total pressure \( \mathcal{P} = \mathcal{P}_s + \mathcal{P}_c \) as functions of density \( \rho \). 
Lines denote values obtained from sedimentation profiles using Eqs.~(\ref{eq:33}) and (\ref{eq:29}) for two gravitational forces \(F_g = -0.01\) (solid) and \(F_g = -0.05\) (dashed), while symbols indicate corresponding values measured in homogeneous periodic simulations. 
For each \(\mathrm{Pe}\), an appropriate gravitational strength \(F_g\) was chosen to ensure well-resolved profiles while preserving the validity of the local density approximation.
Good agreement is observed across all activity levels, confirming the accuracy of the sedimentation-based approach.
The dotted lines correspond to the equation of state of the equilibrium Tonks gas. 
}}
\label{fig:5}
\end{figure*}

Using the expressions summarized in Table~\ref{tab:2}, Fig.~\ref{fig:4}c directly compares pressure components extracted from sedimentation profiles with those measured in homogeneous periodic simulations.  
Curves represent sedimentation-derived values, and points indicate periodic system results.  
We find excellent agreement across all pressure components, demonstrating that the sedimentation profile encodes the same EoS as that measured in bulk systems.  
Importantly, both gravitational strengths give sedimentation profiles and calculated pressures consistent with those computed in the periodic system.
However, for the weaker gravitational force, the accessible density range is narrower due to limited stratification. 
For the stronger field, minor deviations emerge at high density, likely reflecting the breakdown of the LDA near the wall and sedimentation layer.

These results highlight a key practical consideration: recovering the EoS from sedimentation requires balancing competing effects.  
If \(F_g\) is too weak, the density profile is nearly uniform, yielding limited pressure variation.  
If \(F_g\) is too strong, sharp density gradients can violate the LDA and introduce higher-order stress contributions.  
While such effects could be treated by including gradient corrections,~\cite{Omar2023-kn,lebowitz1963integral,yang1976molecular,lovett1976structure} we focus on regimes where the LDA is valid.  
We also restrict our analysis to conditions where top wall accumulation is negligible, ensuring the maximum density range is accessible.  
These considerations guide our choice of \(\mathrm{Pe}\) and \(F_g\) values throughout the study.

We now conclude by applying the sedimentation-based approach to extract the pressure EoS across a broad range of Péclet numbers. 
Figure~\ref{fig:5} shows the reduced swim pressure \( \mathcal{P}_s \), collisional pressure \( \mathcal{P}_c \), and total pressure \( \mathcal{P} = \mathcal{P}_s + \mathcal{P}_c \) as functions of density \( \rho \), each computed from sedimentation profiles using the previously outlined procedure and compared against values measured in periodic simulations. 
Curves represent sedimentation-derived values, while symbols denote direct measurements in homogeneous periodic systems.

We vary the gravitational force depending on activity to balance sedimentation profile resolution with the assumptions underlying the local density approximation (LDA). 
For small Péclet numbers (\(\mathrm{Pe} < 0.5 \)), a weaker gravitational field (\(F_g = -0.01\)) avoids substantial bottom wall accumulation and provides a broad range of accessible densities (solid lines). 
At higher activity levels (\(\mathrm{Pe} \geq 0.5 \)),  where the increased persistence of particle motion leads to a broader density profile and greater accumulation near the top wall, a stronger gravitational field (\(F_g = -0.05\)) is used to ensure sufficient stratification (dashed lines).
Across all activity levels, we observe excellent agreement between sedimentation-derived and periodic simulation results.

\section*{Conclusions}

In this study, we outline a theoretical and computational framework for determining an active suspension's pressure equation of state (EoS) from its steady-state sedimentation profile.  
Focusing on the 1D active Brownian particle (1D-ABP) model with purely repulsive interactions, we demonstrated that the total pressure at any height can be recovered from the local number density and polar order profiles via a mechanical force balance. 
This approach accurately reproduces EoS results obtained from simulations with periodic boundary conditions.
A central outcome is the explicit connection between sedimentation data and the different pressure components in active systems.  
The key insight is that the swim pressure, while often framed as a local quantity, enters the mechanical force balance through the integrated polar order, and its behavior is more consistent with a body force density.  

Looking forward, this framework provides a foundation for extending sedimentation-based techniques to more complex systems.  
The methodology we developed should apply generally to ABPs in any spatial dimension, provided the appropriate polarization and density fields can be resolved.  
Moreover, the framework is bidirectional: if the EoS is known, as is the case for the 1D-ABP model [Eqs.~(\ref{eq:18})–(\ref{eq:20})], solving the force balance can predict the density profile.  
The details of this approach will be discussed in a future study.
Conversely, if the density and polarization profiles are known, the EoS can be obtained through simple integration, as done in this work.  
This duality highlights the robustness and flexibility of the sedimentation approach.
A natural next step is considering systems with more complex interactions, such as attractive forces or anisotropic particles, where phase separation and interfacial effects play a more prominent role.  
In particular, it may be possible to extract the binodal of motility-induced phase separation (MIPS) from sedimentation profiles, as has been done for passive systems undergoing liquid-gas coexistence.  

Another promising direction is generalizing this framework to systems with torque-generating interactions, such as those mediated by hydrodynamic coupling, anisotropic interparticle interactions, or shape anisotropy.  
In these cases, the more complicated orientational dynamics will likely require a generalization of the polar order conservation law [eqn~(\ref{eq:30})].  
Incorporating such effects will bring this approach closer to experimental realizations, which often exhibit richer behavior than the ABP model.
Finally, we note that the force driving the formation of density gradients need not be gravitational.  
External electric or magnetic fields, optical gradients, or phoretic forces could be used in place of gravity, expanding potential experimental implementations.~\cite{demirors2018active,pellicciotta2023colloidal,  moyses2016trochoidal,vutukuri2020light,ibrahim2017multiple,ebbens2018catalytic}  

\section*{Author contributions}

Yunhee Choi: numerical simulations, data analysis, visualization, and writing (original draft, review, and editing). 
Elijah Schiltz-Rouse: numerical simulations, data analysis, visualization, and writing (original draft, review, and editing).
Parvin Bayati: numerical simulations, data analysis, visualization, and writing (original draft, review, and editing).
Stewart A. Mallory: conceptualization, numerical simulations, data analysis, visualization, project administration, and writing (original draft, review, and editing).

\section*{Conflicts of interest}
There are no conflicts to declare.

\section*{Data availability}

The simulations in this study were performed using HOOMD-blue version 4.4.1, which is available at \url{https://hoomd-blue.readthedocs.io/en/v4.4.1/}. The analysis scripts and data necessary to reproduce the results and figures in this manuscript are available upon reasonable request.




\balance

\renewcommand\refname{References}


\bibliographystyle{rsc} 


\providecommand*{\mcitethebibliography}{\thebibliography}
\csname @ifundefined\endcsname{endmcitethebibliography}
{\let\endmcitethebibliography\endthebibliography}{}

\end{document}